\documentclass{sig-alternate}
\usepackage{algorithmic}
\usepackage{algorithm}
\usepackage{subfigure}
\usepackage{graphicx}
\usepackage{epsfig}
\usepackage{cite}
\usepackage{threeparttable}
\usepackage{ntheorem}

\begin{document}

\conferenceinfo{Asia and South Pacific Design Automation Conference (ASPDAC) 2012,} {Jan. 30 -- Feb. 2, 2012, Sydney, Australia.}

%theorem
\theoremstyle{plain}
\theoremheaderfont{\normalfont\bfseries}\theorembodyfont{\slshape}
\theoremsymbol{\ensuremath{\diamondsuit}} \theoremseparator{:}
\newtheorem{mytheorem}{Theorem}

%lemma
%\theoremstyle{changebreak}
\theoremsymbol{\ensuremath{\heartsuit}}
%\theoremindent0.5cm
\theoremnumbering{greek}
\newtheorem{mylemma}{Lemma}

%example
\theoremstyle{change}
\theorembodyfont{\upshape}
\theoremsymbol{\ensuremath{\ast}}
\theoremseparator{}
\newtheorem{myxample}{Example}

%proof
\theoremheaderfont{\sc}\theorembodyfont{\upshape}
\theoremstyle{nonumberplain} \theoremseparator{}
\theoremsymbol{\rule{1ex}{1ex}}
\newtheorem{myproof}{Proof}

%definition
\theoremstyle{plain} \theoremsymbol{\ensuremath{\clubsuit}}
\theoremseparator{.}
\theoremnumbering{arabic}
\newtheorem{mydefinition}{Definition}

%%%%%%%%%%%%%%%%%%%%%%%%%%%% Setting to control figure placement
% These determine the rules used to place floating objects like figures
% They are only guides, but read the manual to see the effect of each.
%\renewcommand{\topfraction}{1.9}
%\renewcommand{\bottomfraction}{1.9}
%\renewcommand{\textfraction}{.1}
%

\pagestyle{empty} %% disallow page numbers

\title{EPIC: Efficient Prediction of IC Manufacturing Hotspots With a Unified Meta-Classification Formulation
}

\author{
\authorblockN{
Duo Ding, Bei Yu, Ghosh Joydeep
 and
David Z. Pan\\}
\authorblockA{Dept. of ECE, The University of Texas at Austin, Austin, TX 78712\\}
\authorblockA{\authorrefmark{1}IBM T. J. Watson Research Center, Yorktown Heights, NY 10598\\}
\{ ding, dpan \}@cerc.utexas.edu}
\numberofauthors{1}
\author{
\alignauthor Duo Ding, Bei Yu, Joydeep Ghosh and David Z. Pan\\
\affaddr{ECE Dept. Univ. of Texas at Austin, Austin, TX 78712} \\
%\affaddr{{\large{$^\ast$}}IBM T. J. Watson Research Center, Yorktown Heights, NY 10598} \\
\affaddr{\{ding, bei\}@cerc.utexas.edu, \{ghosh, dpan\}@ece.utexas.edu}
}

%\small
{

\maketitle
\thispagestyle{empty}%essential

\begin{abstract}
In this paper we present \emph{EPIC}, an efficient and effective predictor for IC
manufacturing hotspots in deep sub-wavelength lithography.
EPIC proposes a unified framework to combine different hotspot detection methods
together, such as machine learning and pattern matching, using mathematical
programming/optimization. EPIC algorithm has been tested  on a number of industry
benchmarks under advanced manufacturing conditions. It demonstrates so far the best
capability in selectively combining the desirable features of various hotspot
detection methods (3.5-8.2\% accuracy improvement) as well as significant
suppression of the detection noise (e.g., 80\% false-alarm reduction). These
characteristics make EPIC very suitable for conducting high performance physical
verification and guiding efficient manufacturability friendly physical design. 
\end{abstract}
\renewcommand{\thefootnote}{}
\footnote{\scriptsize This work is supported in part by NSF, SRC, Oracle and NSFC.}
\footnote{\scriptsize Asia and South Pacific Design Automation Conference (ASPDAC), Jan. 30 -- Feb. 2 2012, Sydney, Australia.}

%
%\vspace{-0.05in}
%\category{B.7.2}{Hardware, Integrated Circuit} Design Aids
%\vspace{-0.15in}\terms{Algorithms, Design, Performance}
%\vspace{-0.15in}\keywords{VLSI, Track Routing, Yield, Manufacturability}
%\vspace{0.1in}

%
\category{B.7.2}{Hardware, Integrated Circuit} Design Aids
%
%\terms{Algorithms, Design, Performance}
%
\keywords{Design for Manufacturability, Lithography Hotspots, Meta Classification, Machine Learning, Pattern Matching}

\section{Introduction}
\label{intro}

Due to the widening gap between the continuous scaling of feature-size and the limited lithography capability\cite{itrs08}, the semiconductor industry is critically challenged in both IC design and manufacturing.
To address these challenges, design-aware manufacturing and manufacturing-friendly design techniques have been developed to avoid high variability design patterns (process hotspots) and to ensure high product yield at post Silicon stage.
During such processes, printing masks are usually re-targeted and optimized through powerful resolution enhancement techniques (RETs) such as Sub-Resolution Assist Features, Optical Proximity Correction, etc.
At the same time, various merging lithography technologies are under active research and development, including Double (Multiple) Patterning lithography, E-beam lithography and EUV lithography.
However, these technologies still suffer from different degrees and types of printing variabilities, therefore layout dependent lithography hotspots remain a challenging issue.

To optimize the masks of a design for better printability, one approach is to first locate the lithography hotspots in a layout, then fix them in a construct-by-correction manner.
Meanwhile in recent years, CAD methodologies have evolved to incorporate RET models into early design stages (e.g., detailed routing) to avoid lithography-unfriendly patterns in a correct-by-construction manner~\cite{Mitra05:dac,Minsik08:dac,Tai-Chen08:dac,D.Z.Pan10:FTEDA,Duo11:dac}.
Consequently, fast and accurate lithography hotspot detection becomes a common and critical issue for a wide range of applications in both design and manufacturing.

However, the quests for such detection methods have been critically challenged in many aspects: (1) designs are getting more complex; (2) under the evolving manufacturing conditions, the number of real hotspots is only a very small fraction of the entire design, making it very difficult to achieve high detection accuracies and low false-alarms simultaneously; (3) detections are seriously run-time constrained due to short turn-around-time, etc.

Current state-of-the-art hotspot detection methods mainly fall into 3 categories.
(1) lithography simulations are very accurate but CPU intensive.
(2) Machine learning techniques~\cite{Norimasa07:spie,Duo09:icicdt,Drago09:dac,Duo11:aspdac,Jen-Yi09:spie,Jen-Yi11:aspdac,DuoMLK11:tcad} with good noise suppression capability are still in need of further accuracy improvement.
(3) Pattern matching techniques\cite{Jingyu07:iccad,A.B.K06:spie,Yao06:iccad,Ning08:spie} that are very good at detecting pre-characterized hotspot patterns lack the capability to predict never-before-seen hotspots.
This is especially problematic when new types of designs are involved after the original pattern library is built.

Recently in~\cite{Jen-Yi11:spie}, a hotspot detection flow was proposed to hybridize the strengths of machine learning models and pattern matching models.
Such a flow feeds data samples to a pattern matcher first, then employs machine learning classifiers to further examine the non-hotspot data set produced by the pattern matcher.
It demonstrates good performance trade-off between detection accuracies and false-alarms suppression compared to the previous works.
However, its ad-hoc nature can make the performance fine-tuning and optimization processes very costly.

In order to better address the problem, we propose \emph{EPIC}: an efficient meta-classification formulation (Fig.~\ref{MLK-EPIC:motivation-combo-classifier}) to combine various hotspot detection techniques into a unified and automated framework that selectively adopts their strengths and suppresses their drawbacks. Based on the theoretical framework in Order Statistics \cite{Kangan02:PAA}, we propose a new CAD flow with different types of \emph{base classifiers} and optimize the flow via constrained quadratic programming.

\begin{figure}[b]
%\vspace{-0.15in}
 \centering
\includegraphics[width=3.3in]{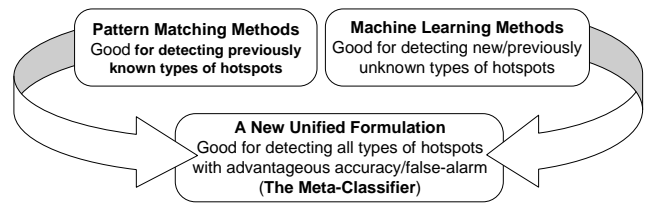}
%\vspace{-0.1in}
  \caption{A new unified formulation for combining various lithography hotspot detection techniques}
  \label{MLK-EPIC:motivation-combo-classifier}
  %\vspace{-0.15in}
\end{figure}

The rest of the paper is organized as follows. In Section~\ref{motiv} we further motivate the \emph{meta-classification} methodology and summarize our main contributions. Section~\ref{Meta-Classifier} details the overall CAD flow together with an overview of the \emph{meta-classifier} construction.
Section~\ref{Base-Classifier} gives a brief description of several classes of building-block detection techniques, followed by Section~\ref{Sec:likelihood}, where these techniques are combined under the \emph{meta-classifier} with mathematical programming and optimization techniques. Section~\ref{EPIC:simu} presents the results and analysis. Section~\ref{conc} concludes the paper.

\begin{figure}[t]
%\vspace{-0.1in}
 \centering
\includegraphics[width=3.3in]{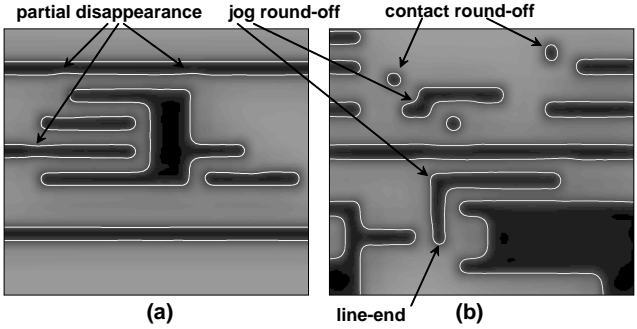}
%\vspace{-0.2in}
  \caption{Examples of lithography hotspot patterns}
  \label{MLK-EPIC:motivation-hotspot2}
  %\vspace{-0.15in}
\end{figure}

\begin{figure*}[t]
%\vspace{-0.05in}
 \centering
\includegraphics[width=7.0in]{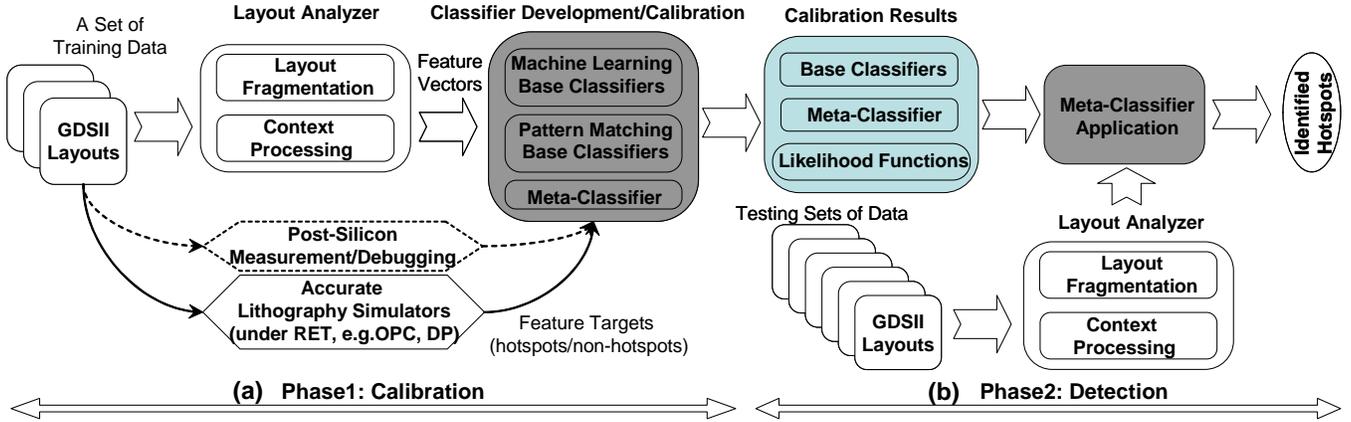}
%\vspace{-0.2in}
  \caption{The overall CAD flow proposed for hotspot detection based on meta-classification formulation}
  \label{MLK-EPIC:meta-flow}
  %\vspace{-0.05in}
\end{figure*}

\section{Motivation and Contributions}
\label{motiv}

In the previous section we discussed the need for a systematic and unified meta-classification methodology to selectively combine certain features of multiple hotspot detection engines. In this section, we use an example to further motivate such a meta-classifier.

Fig.~\ref{MLK-EPIC:motivation-hotspot2} presents the printed images of 2 local regions from certain design at 32$nm$ technology node after applying RETs.
We can make several observations from Fig.~\ref{MLK-EPIC:motivation-hotspot2}(a) and (b).
First, there are various types of process hotspots, featuring complex patterns related to line-ends, jogs, corners or contacts, etc.
Second, hotspot patterns suffer from different amount of manufacturing variation, which is usually measured by the Edge Placement Error (EPE).
Therefore by setting different EPE thresholds, we can classify lithography hotspots into multiple categories.
This allows us more efficiency to study the manufacturability and yield effects of each category.
We first define lithography hotspots:
\begin{mydefinition}
    \textbf{\emph{Hotspot}}: A pattern (or part of a pattern) in an IC design layout that suffers excessive EPE/variation under lithography printing variation at fabrication stage.
\end{mydefinition}
The definition of lithography hotspots is dependent on the EPE error tolerance of a design. Excessive EPE can lead to electrical errors (parasitics variation, timing issues, etc.) or even logic errors (shorts, opens, etc.). To avoid these issues and assist design sign-off and manufacturing closure, lithography hotspots should be properly predicted and avoided during early design stages with short turn-around-time.
In this paper, we use the following two targets to quantitatively calibrate the prediction performance:
\begin{mydefinition}
    \textbf{\emph{Hotspot Accuracy}}: The rate of correctly predicted hotspots among the set of actual hotspots.
\end{mydefinition}
This rate characterizes the success rate of hotspot prediction within the set of actual hotspots. We also use \emph{Hit} to represent the actual count of correctly predicted hotspots, or equivalently, the rate of \emph{Hotspot Accuracy} percentage wise.

\begin{mydefinition}
    \textbf{\emph{Hotspot False-Alarm}}: The rate of incorrectly predicted non-hotspots over the set of actual hotspots.
\end{mydefinition}
This rate represents the over-shoots of the prediction, i.e., the set of non-hotspots predicted incorrectly as hotspots. We also use \emph{Extra} to denote the actual count of such a set, or equivalently, the rate of \emph{Hotspot False-Alarm} in percentage.

Next we motivate a meta-classification flow to concurrently optimize \emph{Hit} and \emph{Extra} on top of powerful hotspot prediction methods.
During our prediction process, each fragment geometry in the layout will be processed and analyzed by multiple hotspot detection engines. Suppose we input pattern $i$ to a machine learning classifier ML and a pattern matcher PM at the same time and the prediction results are $x_i^{ML}$ and $x_i^{PM}$, respectively. $x_i^{ML}$ takes certain value between -1 (non-hotspot) and +1 (hotspot), while $x_i^{PM}$ usually is either -1 (non-hotspot) or +1 (hotspot). Thus the simplified meta-classification problem becomes the following motivational problem:

\textbf{\emph{Given} decisions $x_i^{ML}$ and $x_i^{PM}$ over the same design pattern $i$, \emph{decide} the final hotspot target label $T_i^{meta}$ to simultaneously maximize \emph{Hotspot Accuracy} and minimize \emph{Hotspot False-Alarms}.}

First, it is easy to see that $T_i^{meta}$ is +1 if both $x_i^{ML}$ and $x_i^{PM}$ are (very close to) +1; $T_i^{meta}$ is -1 if both $x_i^{PM}$ and $x_i^{ML}$ are (very close to) -1. Second, in the cases when $x_i^{ML}$ and $x_i^{PM}$ disagree with each other, we introduce the \emph{weighting functions} $f^{ML}(x)$ and $f^{PM}(x)$ to adjust the weights and improve detection performance.
\begin{equation}
    T_i^{meta} = \digamma^{meta}\{ x_i^{ML} \cdot f^{ML}(x_i^{ML}) + x_i^{PM} \cdot f^{PM}(x_i^{PM})\}
    \label{meta-def-motiv}
\end{equation}
If we define $T_i^{meta}$ as in Eqn.(\ref{meta-def-motiv}) above, we can pre-calibrate the \emph{weighting functions} with accurate lithography simulations as golden targets. Then we can use the calibrated functions onto new layout fragments by applying Eqn.(\ref{meta-def-motiv}), where $\digamma^{meta}$ is a threshold cut-off function defined as follows,
\begin{equation}
    \digamma^{meta}(x) = \left\{
                          \begin{array}{ll}
                            +1~(hotspot), & \hbox{if~$x$$\geq$$\theta$} \\
                            -1~(nonhotspot), & \hbox{if~$x$$<$$\theta$}
                          \end{array}
                        \right.
\label{threshold_func}
\end{equation}
Such a formulation combines machine learning and pattern matching techniques meanwhile preserves generality to cover both cases, as if $f^{ML}(x)$=1 and $f^{PM}(x)$=0, then Eqn.(\ref{meta-def-motiv}) degenerates into a machine learning classifier ML; similarly for a pattern matcher PM. Therefore the solution to the above motivational problem lies in the configuration and optimization of the \emph{weighting functions} such that the overall hotspot prediction performance exceeds each individual predictor. Built upon such a motivation, this paper proposes a systematic CAD flow to construct and optimize a \emph{meta-classifier} integrating multiple types of powerful hotspot prediction techniques (known as disparate \emph{base classifiers}).
Our key contributions are as follows,

\begin{itemize}
  \item We propose for the first time a unified \emph{meta-classifier} to seamlessly combine the advantages of various hotspot detection techniques for enhanced accuracy and reduced false-alarms.
  \item We develop high performance hotspot detection engines as \emph{base classifiers} to leverage state-of-the-art machine learning and pattern matching techniques.
  \item We employ Quadratic Programming techniques to achieve efficient configuration and performance optimization of the meta-classifier.
  \item We perform exhaustive assessment on the proposed method using various industry-strength benchmarks under advanced RET and manufacturing conditions.
\end{itemize}

\section{Meta-Classification Overview}
\label{Meta-Classifier}

\subsection{Overall Flow}

Fig.~\ref{MLK-EPIC:meta-flow} shows the overall flow for calibrating and applying the \emph{meta-classifier}.
It consists of 2 steps, the calibration and the detection phases.
Before going into details, we introduce the following key components of the proposed flow.
\begin{mydefinition}
    \textbf{\emph{Base Classifier}}: An individual hotspot classifier that is optimized under certain performance metric, such as detection accuracy, or false-alarms, or adaptivity to new unknown designs, etc.
\end{mydefinition}
\begin{mydefinition}
    \textbf{\emph{Weighting Function}}: A function that properly weights and compensates the prediction result of a \emph{base classifier} such that the overall combinations of individual \emph{base classifiers} can be configured for better accuracy and smaller noise.
\end{mydefinition}
\begin{mydefinition}
    \textbf{\emph{Meta-Classifier}}: A classifier that is formulated and optimized via proper combinations of multiple \emph{base classifiers} under a set of \emph{weighting functions} to further enhance hotspot prediction performance.
\end{mydefinition}

According to Fig.~\ref{MLK-EPIC:meta-flow}, Phase1 is the calibration stage where the \emph{base classifiers} and the \emph{weighting functions} are configured and optimized using training data sets.
This stage requires the supervision of accurate lithography simulators or real silicon debugging data.
Phase2 is the stage when the established \emph{meta-classifier} is applied onto new testing data sets.
This stage operates at very high speed without accurate lithography simulations.

\subsection{Constructing the Meta-Classifier}

The construction and optimization of the \emph{meta-classifier} are the two key contributions of this paper. In this section we give an illustrative overview of the proposed methodology, leaving detailed analysis to Section~\ref{Sec:likelihood}.

The development of a \emph{meta-classifier} is illustrated in Fig.~\ref{MLK-EPIC:meta-machine-core}, which is mainly divided into 3 levels.
For every layout pattern geometry $i$, certain key hotspot features are extracted then fed into each \emph{base classifier}.
\emph{Base classifiers} generate the prediction decisions ($x_i$'s) of pattern $i$, then the weight of each classifier's decision is generated by the \emph{weighting functions}.
The final meta-decision is the weighed sum of \emph{base classifiers}.
Generalizing from the motivational example, we define the following:
\begin{equation}
    T_i^{meta} = \digamma ^ {meta} \{\sum^{N}_{k=1}{x_i^{(k)} \cdot f^{(k)}(x_i^{(k)})}\}
\label{meta-decision}
\end{equation}
where $T_i^{meta}$ is the final decision value of pattern $i$, $N$ is the total number of \emph{base classifiers}, $f^{(k)}(\cdot)$ is the \emph{weighting function} of the $k$th \emph{base classifier}, $x_i^{(k)}$ is the output from the $k$th \emph{base classifier} when pattern $i$ is the input. $\digamma^{meta}$ is the same as in Eqn.(\ref{threshold_func}).

\begin{figure}[h]
%\vspace{-0.15in}
 \centering
\includegraphics[width=3.3in]{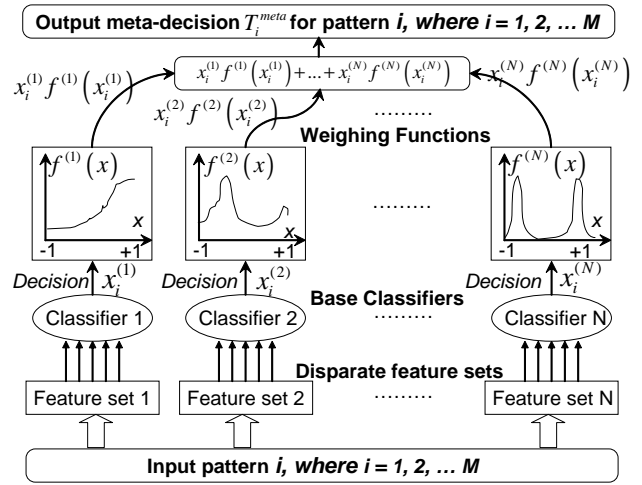}
%\vspace{-0.1in}
  \caption{\emph{Meta-classifier} construction via a combination of disparate \emph{base classifiers}}
  \label{MLK-EPIC:meta-machine-core}
  %\vspace{-0.15in}
\end{figure}

By examining the corner cases, we notice that if the \emph{weighting function} of SVM \emph{base classifier} $f^{SVM}(x)\equiv1$, and all other \emph{base classifiers} have 0 value \emph{weighting functions}, then the \emph{meta-classifier} degenerates into a SVM \emph{base classifier}. Therefore by adjusting the \emph{weighting functions} we can achieve a performance trade-off between different types of hotspot detection techniques, such as machine learning and pattern matching. In the following sections, we will discuss the development and optimization of each classifier involved in the proposed flow.

\section{Constructing Base Classifiers}
\label{Base-Classifier}

In this section, we elaborate the \emph{base classifiers} using machine learning techniques (Artificial Neural Network and Support Vector Machine) and pattern matching techniques.

\subsection{Artificial Neural Network Classifiers}

The ANN \emph{base classifier} is built via fine tuning\cite{Duo11:aspdac,DuoMLK11:tcad}. We briefly describe the formulation and our specific feature contents as follows.
%\begin{scriptsize}
%\vspace{-0.1in}
\begin{equation}
objective : minimize\{\sum_{p=1}^{N} E^{p}\} \quad w.r.t \quad \omega_{ij}, \omega_{jk}
\label{ORouter:equation1}
\end{equation}
\vspace{-0.1in}
\begin{equation}
E^{p} = \frac{1}{2}[out_{p} - y_{p}]^2
\label{ORouter:equation2}
\end{equation}
\vspace{-0.1in}
\begin{equation}
out_{p} = f_{out}\{\sum_{j} \omega_{jk}\cdot f_{hid}(\sum_{i}V_{p}^{i}\cdot \omega_{ij})\}
\label{ORouter:equation3}
\end{equation}
\vspace{-0.2in}
\begin{equation}
\frac {\partial E^{p}} {\partial \omega_{jk}} = (out_{p} - y_{p}) \cdot f_{hid}\{\sum_{i} V_{p}^{i} \cdot \omega_{ij}\}
\label{ORouter:equation4}
\end{equation}
\vspace{-0.2in}
\begin{equation}
\frac {\partial E^{p}} {\partial \omega_{ij}} = (out_{p} - y_{p}) \cdot \omega_{jk} \cdot V_{p}^{i} \cdot (1+ out_{hid}^{j})(1-out_{hid}^{j})
\label{ORouter:equation5}
\end{equation}
\vspace{-0.2in}
\begin{equation}
f_{hid} = \frac{2}{(1+e^{-2x})}-1, \qquad f_{in} = f_{out} = x
\label{ORouter:equation6}
\end{equation}
\vspace{-0.2in}
\begin{equation}
sign\_func (x)=
    \left\{
    \begin{array}{cc}
    -1  &   \text{x $<$ 0} \\
    0  &  \text{x $=$ 0} \\
    +1       &   \text{x $>$ 0}
    \end{array}
    \right.
\label{ORouter:equation6dot5}
\end{equation}
\vspace{-0.05in}
\begin{equation}
Est_{\tilde{p}} = \digamma^{ann}\{f_{out}[\sum_{j} \omega_{jk}\cdot f_{hid}(\sum_{i}V_{\tilde{p}}^{i}\cdot \omega_{ij})]\}
\label{ORouter:equation7}
\end{equation}
%\vspace{-0.1in}
%\end{scriptsize}
An ANN classifier (predictor) works by calculating an outcome $out_p$ for a data sample vector $V_{p}$ based on established weights ($\omega_{jk}$) and biases assigned to a neural network structure, such that the summed square error is minimized according to Eqn.(\ref{ORouter:equation1}) and that $out_p$ approximates certain target $y_p$.
The models shown here are customized with single hidden layer of neurons, with transfer functions denoted as $f_{hid}$. Inputs $V_{p}$ to the ANN kernels are the extracted feature vector samples labeled with values ($y_{p}$) indicating hotspot or nonhotspot patterns (these values can be continuous for variability prediction). We use $p$ to represent feature vector index with $p$ = $1$ to $N$, $V_{p}^{i}$ denotes the $i$th element of vector $V_{p}$, $i$ = $1$ to $M$, where $M$ is the total number of features for each sample vector. We use $f_{in}$ and $f_{out}$ to represent input and output layer transfer functions, and index $i$, $j$, $k$ to indicate neuron indices in the input, hidden and output layer respectively. $\digamma^{ann}$ is the threshold adjustment for performance fine-tuning.
Once the ANN \emph{base classifier} is fully calibrated, we can apply it to estimate $Est_{\tilde{p}}$ according to Eqn.(\ref{ORouter:equation7}) without using costly lithography simulations.

\subsection{Support Vector Machine Classifiers}

Inside the meta-machine block, we employ a $C$-class Support Vector Machine (SVM) classifier fine-tuned based on\cite{Duo11:aspdac,DuoMLK11:tcad}. We brief the problem formulation as follows.
%\begin{scriptsize}
\begin{equation}
objective : minimize\{f(\alpha) = \frac{1}{2} \alpha ^{T}Z \alpha - e^{T} \alpha\} \quad w.r.t \quad \alpha
\label{Hot:equation8}
\end{equation}
\vspace{-0.2in}
\begin{equation}
subject \quad to: 0 \leq \alpha_{i} \leq C, ~~i=1,...,n
\label{Hot:equation9}
\end{equation}
\vspace{-0.2in}
\begin{equation}
y^{T} \cdot \alpha = 0
\label{Hot:equation10}
\end{equation}
\vspace{-0.2in}
\begin{equation}
K(V_{i}, V_{j}) = exp \{\gamma \cdot \|V_{i} - V_{j}\|^{2}\}
\label{Hot:equation11}
\end{equation}
\vspace{-0.2in}
\begin{equation}
slope\_func (x)=
    \left\{
    \begin{array}{cc}
    0  &   \text{$x$ $\leq$ 0} \\
    x  &  \qquad \text{0 $<$ $x$ $<$ C} \\
    C       &   \text{$x$ $\geq$ C}
    \end{array}
    \right.
\label{Hot:equation12}
\end{equation}
\vspace{-0.1in}
\begin{equation}
Est_{\tilde{p}} = \digamma^{svm}\{\sum_{i} {\alpha_{i} y_{i} K(V_{\tilde{p}}, V_{i}) + bias}\}
\label{Hot:equation13}
\end{equation}
%\vspace{-0.2in}
%\end{scriptsize}
Given $V_{i}$, $i$=$1$ to $M$ sample vectors with $n$ number of features, with label $y_{i}$ (either \emph{hotspot} or \emph{non-hotspot} for 2-class SVM). $e$ is a vector of all 1's. $C$ is a pre-set upper bound to constrain feasible regions for hotspot detection under real manufacturing conditions. $Z$ is $n$ by $n$ positive semi-definite matrix defined as $Z_{ij}$ = $y_{i}y_{j}K(V_{i}, V_{j})$, where $K(V_{i}, V_{j})$ is defined in Eqn.(\ref{Hot:equation11}) as the kernel function. $\alpha$ is the $N$ element weight vector for $V_{p}$'s. $\digamma^{svm}$ is a threshold function to adjust and fine-tune the estimation performance of $Est_{\tilde{p}}$.

The configuration of SVM \emph{base classifiers} is achieved through performing a set of algorithms over the calibration data $V_p$'s to identify the support vectors and weight coefficients that construct a classification hyper-plane with maximized separation margin.
Once configured, we apply the SVM model to evaluate new data samples according to Eqn.(\ref{Hot:equation13}) without costly lithography simulations.

\subsection{Pattern Matching Classifiers}

\begin{figure}[h]
\vspace{-0.05in}
 \centering
\includegraphics[width=3.3in]{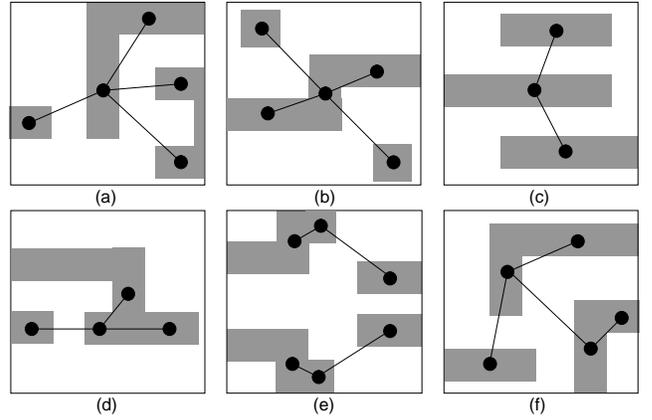}
\vspace{-0.3in}
  \caption{Example patterns in PM \emph{base classifiers}}
  \label{MLK-EPIC:pattern-matching}
 \vspace{-0.15in}
\end{figure}

We explored the current state-of-the-art methods~\cite{A.B.K06:spie, Yao06:iccad, Jingyu07:iccad} and came up with several major classes (each with hundreds of specific hotspot structures) of pattern matching \emph{base classifiers} to cover various types of lithography hotspots, relating to special line-ends, corners, jogs, contact patterns, etc.

Some example hotspot patterns are illustrated in Fig.~\ref{MLK-EPIC:pattern-matching}. In particular, we have fine-tuned the pattern matchers to have broader pattern coverage rather than performing exact matching. As a result, the established pattern matchers demonstrate very good hotspot accuracies onto new data sets. Obviously, the penalty of such fine-tuning is the consequent high false-alarms. However, as we will see later in Section~\ref{EPIC:simu}, the \emph{meta-classifier} performs well in suppressing the false-alarms of such a PM \emph{base classifier}.

\begin{figure}[t]
%\vspace{-0.05in}
 \centering
\includegraphics[width=3.3in]{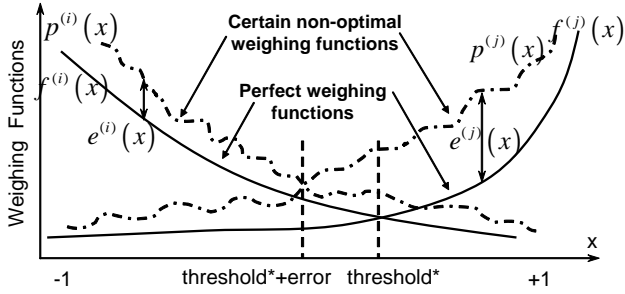}
\vspace{-0.3in}
  \caption{An illustration of \emph{weighting function} error analysis, assuming two \emph{base classifiers} (N=2)}
  \label{MLK-EPIC:meta-machine-error}
  \vspace{-0.15in}
\end{figure}

\section{Optimizing Meta-Classification}
\label{Sec:likelihood}

Given the proposed \emph{meta-classifier} in Fig.~\ref{MLK-EPIC:meta-machine-core}, in this section we first analyze the Mean-Square-Error of the \emph{meta-classifier} introduced by the errors/noises of the \emph{weighting functions}. Then we propose mathematical programming techniques to optimize the \emph{weighting functions} to minimize the detection error.

\subsection{Meta-Classification Error Analysis}

Depicted in Fig.~\ref{MLK-EPIC:meta-machine-error} are 2 sets of curves. Assume the black curves are the optimal \emph{weighting functions} and the intersected point \emph{threshold*} is the optimal cutoff value for $\digamma ^{meta}(\cdot)$ (this happens under 1:1 importance ratio between \emph{hotspot accuracy} improvement and \emph{hotspot false-alarm} reduction). Suppose the dotted curves are the sub-optimal \emph{weighting functions} for the $i$th and $j$th \emph{base classifiers}. In this case, the derived cut-off threshold becomes \emph{threshold*+error}, thus the meta-classification flow has an error:
\begin{equation}
    MSE^{noise} = \int_{x}\{ \sum^{N}_{k=1}{f^{(k)}(x) \cdot x - \sum^{N}_{k=1}{p^{(k)}(x) \cdot x}} \}^{2} dx
\label{error:1}
\end{equation}
\vspace{-0.1in}
\begin{equation}
    ~ = \sum^{N}_{k=1}{\int_{x} \{ [f^{(k)}(x) - p^{(k)}(x)] \cdot x \}^2 dx}
\label{error:2}
\end{equation}
\vspace{-0.1in}
\begin{equation}
    ~ = \sum^{N}_{k=1}{\int_{x} \{ [e^{(k)}(x)] \cdot x \}^2 dx}
\label{error:3}
\end{equation}
%\vspace{-0.1in}
\begin{equation}
    e^{(k)}(x) = f^{(k)}(x) - p^{(k)}(x)
\label{error:5}
\end{equation}

From the analysis above, we observe that the classification error accumulates among all \emph{base classifiers} with a quadratic index on each term, should noise/error occur in the \emph{weighting functions}. Therefore, it is critical to find the optimal \emph{weighting functions} to ensure the \emph{meta-classifier}'s noise robustness.
In the following section, we will explore the mathematical formulation that optimizes the \emph{weighting functions} given certain calibration data.

\subsection{Weighting Function Optimization}
\label{Mapping_Func}

We first define the \emph{meta-classification} Mean-Square-Error over the entire calibration data set under the supervision of accurate lithography simulations:
\begin{equation}
    MSE^{meta} = \frac{1}{M} \cdot \sum^{M}_{i=1}{ \|\sum^{N}_{k=1}{x_i^{(k)} \cdot p^{(k)}(x_i^{(k)})}-T^{litho}_{i}\|^2}
\end{equation}
where $M$ is the total number of calibration samples and $T^{litho}_i$ is the baseline hotspot characterization result given by accurate lithography simulator. Table~\ref{EPIC2011:VarTerms} details the short-hand terms used in our mathematical formulation.

To minimize the Mean-Square-Error among the sample space meanwhile avoid over-fitting of the training data set, we define the performance optimization formulation:
\begin{equation}
To~ minimize:~ MSE^{meta}~ +~ PCost~~~w.r.t~p^{(k)}(x_i^{(k)})
\label{equ:orig_formu1}
\end{equation}
\begin{equation}
     PCost = \lambda_0\sum_{i}{\sum_{k}{(p^{(k)}(x_i^{(k)})-const)^2}}
\end{equation}
where $\lambda_0$ is a non-negative penalty applied to constrain the calibration process such that the \emph{weighting functions} are bounded within certain proximity of a constant parameter. This will prevent numerical instability and preserve detection generality of the \emph{weighting functions} when applied to new testing data. Such proximity is adjustable by varying $\lambda_0$.

To assist numerical optimization, we quantized the original continuous \emph{weighting functions} $p^{(1)}(x)$ $\sim$ $p^{(N)}(x)$, each into $L(k)$ levels, with each level being a single weight value denoted as $p_k^{(l)}$, where $l$ $\in$ [1,$L(k)$].

%%%%%%%%%%%%%%%%%%%%%%%%%%%%%%%%%%%%%%%%%%%%%%%%%%%%%%%%%%%%%%%%%%%%%%%%%%%%%%%%%%%%%%%
\begin{table} [t]
%\vspace{-0.15in}
\centering

\caption{Variables and terms in the QP formulation}
\label{EPIC2011:VarTerms}
\begin{scriptsize}
\begin{tabular}{||c|c||}
\hline
\hline
 Terms & Descriptions \\
\hline
\hline
        $N$ & Number of the \emph{base classifiers}\\
\hline
        $M$ & Number of \emph{meta-machine} calibration sample data\\
\hline
        $i$ & Index of each input sample pattern\\
\hline
        $k$ & Index of each \emph{base classifier}\\
\hline
         & Prediction result from the \emph{base classifier} $k$\\
        $x_{i}^{(k)}$ & given input data sample $i$\\
\hline
        $f^{(k)}(x_i^{(k)})$ & Value of perfect \emph{weighting func.} $f^{(k)}(\cdot)$ at $x_{i}^{(k)}$\\
\hline
        $p^{(k)}(x_i^{(k)})$ & Value of non-perfect \emph{weighting func.} $p^{(k)}(\cdot)$ at $x_{i}^{(k)}$\\
\hline
        $L^{(k)}$ & Total quantization levels of \emph{base classifier} $k$\\
\hline
        $l$ & Index of each quantization level\\
\hline
         & Quantized weight value from $f^{(k)}(\cdot)$\\
        $p_k^{(l)}$ & at level $l$ of the \emph{base classifier} $k$, $p_k^{l}$ $\in$ $[1,~L^{(k)}]$\\
\hline
        & The quantization mapping function: \\
       $\Theta(\cdot)$ & $x_i^{(k)}$ $\rightarrow$ quantization level index $l$\\
\hline
        & Prediction result given by \emph{base classifier} $k$ at\\
        $\alpha_{k}^{(l)}(i)$& level $l$ with input sample $i$ (set to 0 if NULL)\\
        & i.e., the value to which $p_k^{(l)}$ is to be applied\\
\hline
        $L^{total}$ & Total number of independent $p_k^{(l)}$\\
\hline
        $Q$ & A definite positive matrix $\in$ $\Re^{L^{total} \times L^{total}}$\\
\hline
        $c$ & A vector $\in$ $\Re^{L^{total} \times 1}$ \\
\hline
         & Variable vector for the \\
         $X$ & quadratic programming formulation, where\\
        & $X$ = $[p_1^{(1)}~...p_k^{(l)}~...p_N^{(L^{(N)})}]^T$ $\in$ $\Re^{L^{total} \times 1}$\\
\hline
        $T_i^{meta}$ & \emph{Meta-machine} prediction result for input sample $i$ \\
\hline
         & Prediction baseline for input sample $i$ \\
        $T_i^{litho}$ & by accurate lithography simulator \\
\hline
        $\lambda_0$ & Parameter to avoid over-fitting/instability\\
\hline
\hline
\end{tabular}
\end{scriptsize}
%\vspace{-0.1in}
\end{table}
%%%%%%%%%%%%%%%%%%%%%%%%%%%%%%%%%%%%%%%%%%%%%%%%%%%%%%%%%%%%%%%%%%%%%%%%%%

After the \emph{weighting function} quantization process, we have the following modified formulation:
\begin{equation}
To~ minimize:~ \overline{MSE}~ +~ \overline{PCost}~~~w.r.t~p_k^{(l)}
\end{equation}
\begin{equation}
    \overline{MSE} = \frac{1}{M}\sum^M_{i=1}\|\sum_{k=1}^{N}{p_k^{(\Theta(x_i^{(k)}))} \cdot x_i^{(k)}} - T_i^{litho}\|^2
\end{equation}
\begin{equation}
     \overline{PCost} = \lambda_0\sum_{k=1}^{N}{\sum_{l=1}^{L^{(k)}}{(p_k^{(l)}-1)^2}}
\end{equation}
where $p_k^{(l)}$'s are the optimization variables (quantized weight values) for fine-tuning the overall classification quality. In $\overline{PCost}$, we set the constant parameter to 1.0 since it is the boundary factor of numerical up-scaling and down-scaling. Due to the $\overline{PCost}$ term, each weight variable will be scattered not far away from 1.0 meanwhile be optimized under predication error minimization objective. This benefits us in two ways: first, avoiding close to zero weights for the calibration data yields better classification generality over testing data; second, avoiding large weights can maintain good balance among hotspot features meanwhile prevent numeric instability over testing data. For further details of the notations please refer to Table~\ref{EPIC2011:VarTerms}.

Finally we can write the following quadratic programming problem formulation:
\begin{equation}
    f(x) = \frac{1}{2}X^TQX + c^TX
\label{QP_formulate:1}
\end{equation}
\begin{equation}
    X \geq lb
\label{QP_formulate:2}
\end{equation}
\begin{equation}
    lb = [0~ 0~ 0~ 0~ 0~ ...~ 0]^T \in \Re^{L^{total} \times 1}
\end{equation}
where X is the optimization variable vector defined as follows,
\begin{equation}
    X = [p_1^{(1)}~...~p_1^{(L^{(1)})}~...~p_{k}^{(l)}~...~p_N^{(L^{(N)})}]^T \in \Re^{L^{total} \times 1}
\end{equation}
where $L^{total}$ is the total number of $p_k^{(l)}$'s:
\begin{equation}
    L^{total} = \sum^{N}_{k=1}{L^{(k)}}
\end{equation}
Matrix $Q$ is defined as follows,

$Q~=~$
\begin{equation}
    \left(
          \begin{array}{cccccc}
            \beta_{1}^{(1)}{(i)} & \gamma_{1,1}^{(1,2)}{(i)} & . & \gamma_{1,k}^{(1,l)}{(i)} & \gamma_{1,N}^{(1,L^{(N)})}{(i)}\\
            \gamma_{1,1}^{(2,1)}{(i)} & \ddots & . & . & .\\
            . & . & \beta_{1}^{(L^{(1)})}{(i)} & . & .\\
            \gamma_{k,1}^{(l,1)}{(i)} & . & . & \beta_{k}^{(l)}{(i)} & .\\
            \gamma_{N,1}^{(L^{(N)},1)}{(i)} & . & . & . & \beta_{N}^{(L^{(N)})}{(i)}\\
          \end{array}
        \right)
\end{equation}

Vector $c$ is defined as the linear term coefficients vector from the quadratic formulation objective:
\begin{equation}
c = [\omega_{1}^{1}(i)\cdots\omega_{1}^{L^{(1)}}(i)\cdots\omega_{k}^{(l)}(i)\cdots\omega_{N}^{L^{(N)}}(i)~]^T
\end{equation}
where the related terms are defined as follows, and $\alpha_{k}^{(l)}(i)$ is an intermediate term to link the \emph{base classifier}'s prediction values with $p_k^{(l)}$, i.e., $\alpha_{k}^{(l)}(i)$ is the $x_i^{(k)}$ value that falls into level $l$ of the quantized weighting function relating to the \emph{base classifier} $k$. Given certain $i$ and $k$, if there is no output values corresponding to level $l$, then $\alpha_{k}^{(l)}(i)$ is set to 0.
\begin{equation}
\vspace{-0.1in}
    \beta_{k}^{(l)}{(i)} =  \frac{2}{M}{\sum^{M}_{i=1}{[\alpha^{(l)}_{k}{(i)}]^2}} + 2\lambda_0
\end{equation}
\begin{equation}
\vspace{-0.1in}
    \gamma_{m,k}^{(n,l)}{(i)} = \frac{2}{M}\sum_{i=1}^{M}{\alpha_{m}^{(n)}{(i)} \cdot \alpha_{k}^{(l)}{(i)}}
\end{equation}
\begin{equation}
    \omega_{k}^{(l)}(i) = -\frac{2}{M} \sum_{i=1}^{M}{T_i^{litho} \cdot \alpha_{k}^{(l)}(i)} -2\lambda_0
\end{equation}

The values and parameters in the above equations are derived properly so that the original problem in Eqn.(\ref{equ:orig_formu1}) becomes the minimization of a quadratic problem in Eqn.(\ref{QP_formulate:1}). Once it is solved, we apply the resulting \emph{weighting functions} over some calibration data to properly select a threshold function $\digamma^{meta}(\cdot)$ to further balance \emph{hotspot accuracy} and \emph{hotspot false-alarm}. After calibration, the \emph{meta-classifier} will be tested over new design layouts based on Eqn.(\ref{meta-decision}).

\subsection{Complexity Analysis}
\begin{mytheorem}
    Matrix Q is positive definite under certain conditions of $\lambda_0$ and the formulated quadratic programming problem has the following properties: (1) it can be solved in polynomial time complexity; (2) if it has a local minimum, then this local minimum is also the global minimum.
\end{mytheorem}
\begin{myproof}
    For notation simplicity, we assume a vector $\vec{X}$ $\in$ $\Re^{L^{total} \times 1}$ and $X$ $\geq$ $\vec{0}$. Let $\rho_i$ be the coefficient of $\chi_i$, where $\chi_i$ is the $i$th element of $\vec{X}$. Let $L^{total}$ be the total number of quantization levels among all \emph{base classifiers}. Therefore we have the following:
\begin{equation}
    \vec{X}^T Q \vec{X} = \frac{1}{M} \sum^{M}\sum_{j=i+1}^{L^{total}}{\sum_{i=1}^{L^{total}}{( \rho_i \chi_i + \rho_j \chi_j  )^2}} + \Delta
\end{equation}
where
%\vspace{-0.2in}
\begin{equation}
\Delta = - \frac{L^{total}-2}{M} \sum^{M} \sum_{i}^{L^{total}}{\rho_i^2 \chi_i^2} + \lambda_0 \sum_{i}^{L^{total}}{\chi_i^2}
\end{equation}
We can adjust $\lambda_0$ to achieve positive $\Delta$ (in practice usually a $\lambda_0$ slightly larger than $\frac{L^{total}-2}{M} \sum^{M}$).
Therefore $\vec{X}^T Q \vec{X}$ is always positive given non-zero $\vec{X}$. Thus Q is a positive definite matrix and has no negative eigenvalues under the specified condition. Numerical simulations further validate this proof by showing all positive eigenvalues for matrix Q.
Consequently, the Quadratic Programming problem can be solved by the \emph{ellipsoid method}\cite{Ellip:79} in polynomial time.

Since Q is a symmetric positive-definite matrix, $f(\cdot)$ is now a convex function. Thus the quadratic program has a global minimizer if there exists some feasible vector $X^n$ satisfying the constraints and if $f(\cdot)$ is bounded below on the feasible region ($X^n \in \Re^{n}_{+}$). Therefore in search of a local minimum, if found, will guarantee the optimal global minimum.
\end{myproof}

We solve the quadratic programming problem using a proper $\lambda_0$ to optimize the \emph{weighting functions} during the calibration phase. Then we use a heuristic approach to search for the optimal $\digamma^{meta}(\cdot)$ function.
In Algorithm~\ref{Hotspot:pseudo-code-MM-train} and Algorithm~\ref{Hotspot:pseudo-code-MM-apply}, we show the details for the calibration and application of the \emph{meta-classifier}.

%%%%%%%%%%%%%%%%%%%%%%%%Top Level Algorithm for SVM starts here%%%%%%%%%%%%%%%%%%%%%%
\begin{scriptsize}
\begin{algorithm}[t]
\caption{\textbf{\emph{Meta-Classifier}-Calibration}}
 \label{Hotspot:pseudo-code-MM-train}
\begin{algorithmic}
%\begin{scriptsize}
 \REQUIRE {data sample vectors and over-fit penalty $\lambda_0$}
 \STATE{\textbf{Initialize} $Q$, $c$, $\beta_{k}^{(l)}{(i)}$, $\gamma_{k}^{(l)}{(i)}$, $\omega_{k}^{(l)}(i)$}
 \STATE{\textbf{Build} Hierarchical \emph{MLK-ANN}}
 \STATE{\textbf{Build} Hierarchical \emph{MLK-SVM}}
 \STATE{\textbf{Build} Pattern Matchers}

 \FOR {All input data samples}
    \STATE{\textbf{Generate} the \emph{base classifiers}}
    \STATE{\textbf{Update} $Q$, $c$, $\beta_{k}^{(l)}{(i)}$, $\gamma_{k}^{(l)}{(i)}$, $\omega_{k}^{(l)}(i)$}
 \ENDFOR

 \STATE{\textbf{Formulate} Quadratic Programming Problem}
 \IF {$Q$ not positive definite}
    \STATE{\textbf{Increase} calibration data volume}
    \STATE{\textbf{Improve} hotspot feature quality}
    \STATE{\textbf{Adjust} parameter $\lambda_0$}
    \STATE{\textbf{Consider} matrix pre-conditioning}
 \ENDIF

 \STATE{\textbf{Solve} the Quadratic Programming Problem}

 \STATE{\textbf{Optimize} the detection threshold function $\digamma^{meta}(\cdot)$}

\RETURN{\emph{weighting functions} $p_k^{(l)}$ and $\digamma^{meta}(\cdot)$}
%\end{scriptsize}
\end{algorithmic}
\end{algorithm}
\end{scriptsize}
%%%%%%%%%%%%%%%%%%%%%%%%%%%%%%%%%%%%%%%%%%%%%%%%%%%%%%%%%%%%%%%%%%%%%%%%%%%%%%%%%%%%%%%%%%
%%%%%%%%%%%%%%%%%%%%%%%%Top Level Algorithm for SVM starts here%%%%%%%%%%%%%%%%%%%%%%
\begin{scriptsize}
\begin{algorithm}[t]
\caption{\textbf{\emph{Meta-Classier}-Prediction}}
 \label{Hotspot:pseudo-code-MM-apply}
\begin{algorithmic}
%\begin{scriptsize}
 \REQUIRE {data sample vectors}
 \STATE{\textbf{Load} \emph{weighting functions} $p_k^{(l)}$ and $\digamma^{meta}(\cdot)$}
 \STATE{\textbf{Load} all \emph{base classifiers}}
 \STATE{\textbf{Generate} vector $\overrightarrow{x_i}$ from \emph{base classifiers} outputs}
 \FOR {Each data vector $\overrightarrow{x_i}$}
    \STATE{\textbf{Calculate} $T^{meta}_i$=$\digamma^{meta}(\sum_{k=1}^{N}{p_k^{(\Theta(x_i^{(k)}))} \cdot x_i^{(k)}})$}
 \ENDFOR

\RETURN{Meta-decision $\{T_i^{meta}\}$}
%\end{scriptsize}
\end{algorithmic}
\end{algorithm}
\end{scriptsize}
%%%%%%%%%%%%%%%%%%%%%%%%%%%%%%%%%%%%%%%%%%%%%%%%%%%%%%%%%%%%%%%%%%%%%%%%%%%%%%%%%%%%%%%%%%

\section{Simulation and Testing}
\label{EPIC:simu}

\subsection{Benchmarks and Simulation Setups}

To fully evaluate \emph{EPIC}, we employed a number of training data sets and 3 new testing circuit benchmarks in 32nm. These testing circuits include new hotspot patterns that were not present in the training data. We labeled 2 classes of `real' lithographic hotspots based on 2 EPE thresholds. In Table~\ref{MLK-ELIAD:benchmark-table}, C0 is the class0 hotspot patterns whose printed images suffer from above 6nm of EPE; C1 refers to the patterns whose printed images have EPE from 4.5nm to 6nm. Further details of the 3 testing benchmarks are in Table~\ref{MLK-ELIAD:benchmark-table}. To properly evaluate the proposed methods, we perform accurate lithographic simulations as baseline to identify the actual hotspots under industry-strength RETs.

In our simulation, \emph{EPIC} incorporates two types of machine learning methods based on~\cite{Duo11:aspdac,DuoMLK11:tcad} and several pattern matching techniques based on~\cite{A.B.K06:spie, Yao06:iccad, Jingyu07:iccad, Ning08:spie}.
We implement \emph{EPIC} in C++ on 3.2GHz quad-core Linux workstations.

\subsection{Result Analysis and Comparison}

After the quadratic programming problem is solved, we properly select the decision threshold function $\digamma^{meta}(\cdot)$ using some calibration data to balance between \emph{hotspot accuracy} and \emph{hotspot false-alarm}.
To illustrate such performance trade-off, we test \emph{EPIC}, \emph{ANN} and \emph{SVM} over C0 data with a set of varying thresholds and plot the results in Fig.~\ref{MLK-EPIC:final-chart}. We also plot the performance region of the employed pattern matchers, which include up to 4 major classes of hotspot patterns. As we enrich the pattern library gradually with up to more than hundreds of specific patterns and structures, the overall performance becomes a trade-off between enhancing detection accuracy and suppressing false-alarms, especially when there are new unseen types of hotspots in the testing data.

From Fig.~\ref{MLK-EPIC:final-chart} we observe that in the region of above 70\% accuracy, \emph{EPIC} shows higher \emph{hotspot accuracy} than other methods with very similar \emph{hotspot false-alarm}, it also achieves lower \emph{hotspot false-alarm} given similar \emph{hotspot accuracy}.
We also see that pattern matching methods are not good at detecting new types of hotspots without obvious penalty in \emph{hotspot false-alarm}.
In this sense, machine learning can make pattern matching more robust to predict new/unknown hotspots, especially when pattern enumeration becomes costly.

\begin{figure}[t]
%\vspace{-0.2in}
 \centering
\includegraphics[width=3.5in]{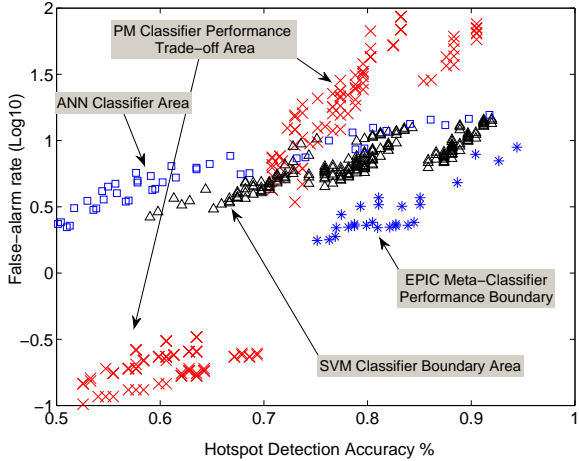}
\vspace{-0.3in}
  \caption{\scriptsize{Trade-off capabilities between hotspot accuracy and false-alarms using various methods on C0 hotspots}}
  \label{MLK-EPIC:final-chart}
  \vspace{-0.3in}
\end{figure}

Based on Fig.~\ref{MLK-EPIC:final-chart}, we calculate the following for each method:
\begin{equation}
\Psi = \alpha \cdot Accuracy^{hotspot} + \beta \cdot False\_alarm^{hotspot}
\end{equation}
where $\alpha$ (positive) and $\beta$ (negative) are user defined parameters to quantify the importance ratio between \emph{hotspot accuracy} and \emph{hotspot false-alarm}. In Table~\ref{MLK-EPIC:benchmark-final-results} and Table~\ref{MLK-EPIC:benchmark-final-comp}, we report the detection result of each method corresponding to the peak of their respective $\Psi$ function.
We observe that \emph{EPIC} reaches the highest performance over both C0 and C1 categories of hotspots in both \emph{hotspot accuracy} and \emph{hotspot false-alarm}.
To be specific, it improves ANN and SVM by 3.5-7\% in \emph{hotspot accuracy} and reduces up to 50\% in \emph{hotspot false-alarm} counts. \emph{EPIC} also outperforms PM by 4.5-8.2\% in \emph{hotspot accuracy} and 53-81\% in \emph{hotspot false-alarm} reduction.
This demonstrates very promising potential of the \emph{meta-classification} flow with respect to \emph{weighting function} optimizations. Moreover, \emph{EPIC} runs at the speed of around 45 min per $mm^2$ design on a 3.2GHz quad-core workstation, which is typically hundreds of times faster than accurate lithography simulator.

%%%%%%%%%%%%%%%%%%%%%%%%%%%%%%%%%%%%%%%%%%%%%%%%%%%%%%%%%%%%%%%%%%%%%%%%%%%%%%%%%
\begin{table} [t]
\centering
\caption{Circuit benchmarks for testing \emph{EPIC}}
\label{MLK-ELIAD:benchmark-table}
\begin{threeparttable}
\begin{small}
\begin{tabular}{|l|c|c|c||}
\hline
\hline
\multicolumn{ 1}{|l|}{Benchmarks} & \multicolumn{ 1}{c|}{CK1} & \multicolumn{ 1}{c|}{CK2} & \multicolumn{ 1}{c|}{CK3}\\
\hline
\hline
\multicolumn{ 1}{|l|}{Layout Size $um^2$}  & \multicolumn{ 1}{c|}{100$\times$100} & \multicolumn{ 1}{c|}{150$\times$150} & \multicolumn{ 1}{c|}{800$\times$800}\\
\hline
\multicolumn{ 1}{|l|}{Fragment number}  & \multicolumn{ 1}{c|}{58K} & \multicolumn{ 1}{c|}{94.5K} & \multicolumn{ 1}{c|}{2.5M }\\
\hline
\multicolumn{ 1}{|l|}{Class0\tnote{a} ~Hotspots number}  & \multicolumn{ 1}{c|}{9} & \multicolumn{ 1}{c|}{21} & \multicolumn{ 1}{c|}{ 122 }\\
\hline
\multicolumn{ 1}{|l|}{Class1\tnote{b}~ Hotspots number}  & \multicolumn{ 1}{c|}{61} & \multicolumn{ 1}{c|}{134} & \multicolumn{ 1}{c|}{ 2.8K}\\
\hline
\hline
\end{tabular}
\end{small}
\begin{scriptsize}
\begin{tablenotes}
\item[a] Class0 hotspots: EPE $\geq$ 6$nm$ for 32$nm$ process.
\item[b] Class1 hotspots: 4.5$nm$ $\leq$ EPE $<$ 6$nm$ for 32$nm$ process.
%\item[c] Class2 hotspots: EPE $<$ 5$nm$ for 45$nm$ process.
\end{tablenotes}
\end{scriptsize}
\end{threeparttable}
\vspace{-0.15in}
\end{table}
%\end{scriptsize}
%%%%%%%%%%%%%%%%%%%%%%%%%%%%%%%%%%%%%%%%%%%%%%%%%%%%%%%%%%%%%%%%%%%%%%%%%%%%%%%%%%%

%%%%%%%%%%%%%%%%%%%%%%%%%%%%%%%%%%%%%%%%%%%%%%%%%%%%%%%%%%%%%%%%%%%%%%%%%%%%%%%%%%%
%\begin{scriptsize}
\begin{table} [t]
%\vspace{-0.15in}
\centering

\caption{Performance of hotspot detection methods}
\label{MLK-EPIC:benchmark-final-results}
\begin{threeparttable}
\begin{small}
\begin{tabular}{|cccccccc|}
\hline
\hline
\multicolumn{ 1}{|c|}{Circuits} & \multicolumn{ 1}{|c|}{Class} & \multicolumn{ 1}{|c|}{Perf.} & \multicolumn{ 1}{|c|}{ANN} & \multicolumn{ 1}{|c|}{SVM} & \multicolumn{ 1}{|c|}{PM} & \multicolumn{ 1}{|c|}{\emph{EPIC}}\\
\hline
\hline
\multicolumn{ 1}{|c|}{} & \multicolumn{ 1}{|c|}{} & \multicolumn{ 1}{|c|}{Hit} & \multicolumn{ 1}{|c|}{6} & \multicolumn{ 1}{|c|}{7} &\multicolumn{ 1}{|c|}{7} &\multicolumn{ 1}{|c|}{9}\\
\cline{3-7}
\multicolumn{ 1}{|c|}{} & \multicolumn{ 1}{|c|}{C0} & \multicolumn{ 1}{|c|}{Extra} & \multicolumn{ 1}{|c|}{79} &\multicolumn{ 1}{|c|}{41} &\multicolumn{ 1}{|c|}{280} &\multicolumn{ 1}{|c|}{48}\\
\cline{2-7}

\multicolumn{ 1}{|c|}{} & \multicolumn{ 1}{|c|}{} & \multicolumn{ 1}{|c|}{Hit} & \multicolumn{ 1}{|c|}{52} &\multicolumn{ 1}{|c|}{54} &\multicolumn{ 1}{|c|}{53} &\multicolumn{ 1}{|c|}{57} \\
\cline{3-7}
\multicolumn{ 1}{|c|}{CK1} & \multicolumn{ 1}{|c|}{C1} & \multicolumn{ 1}{|c|}{Extra} & \multicolumn{ 1}{|c|}{0.55K} &\multicolumn{ 1}{|c|}{0.33K} &\multicolumn{ 1}{|c|}{1.5K} &\multicolumn{ 1}{|c|}{0.3K} \\

\hline

%%%%%%%%%%%%%%%%%%%%%%%%%%%%%%%%%%%%%%%%%%%%%%%%%%%%%%%%%%%%%%%%%%%%%%%%%%%%%%%%%%%%%%%%%%%%%%%%%%%%

\multicolumn{ 1}{|c|}{} & \multicolumn{ 1}{|c|}{} & \multicolumn{ 1}{|c|}{Hit} & \multicolumn{ 1}{|c|}{18} & \multicolumn{ 1}{|c|}{17} &\multicolumn{ 1}{|c|}{17} &\multicolumn{ 1}{|c|}{19} \\
\cline{3-7}
\multicolumn{ 1}{|c|}{} & \multicolumn{ 1}{|c|}{C0} & \multicolumn{ 1}{|c|}{Extra} & \multicolumn{ 1}{|c|}{0.2K} &\multicolumn{ 1}{|c|}{0.11K} &\multicolumn{ 1}{|c|}{0.7K} &\multicolumn{ 1}{|c|}{0.1K} \\
\cline{2-7}

\multicolumn{ 1}{|c|}{} & \multicolumn{ 1}{|c|}{} & \multicolumn{ 1}{|c|}{Hit} & \multicolumn{ 1}{|c|}{119} &\multicolumn{ 1}{|c|}{120} &\multicolumn{ 1}{|c|}{120} &\multicolumn{ 1}{|c|}{125} \\
\cline{3-7}
\multicolumn{ 1}{|c|}{CK2} & \multicolumn{ 1}{|c|}{C1} & \multicolumn{ 1}{|c|}{Extra} & \multicolumn{ 1}{|c|}{1.2K} &\multicolumn{ 1}{|c|}{0.75K} &\multicolumn{ 1}{|c|}{3.4K} &\multicolumn{ 1}{|c|}{0.65K} \\
\hline
%%%%%%%%%%%%%%%%%%%%%%%%%%%%%%%%%%%%%%%%%%%%%%%%%%%%%%%%%%%%%%%%%%%%%%%%%%%%%%%%%%%%%%%%%%%%%%%%%%%%

\multicolumn{ 1}{|c|}{} & \multicolumn{ 1}{|c|}{} & \multicolumn{ 1}{|c|}{Hit} & \multicolumn{ 1}{|c|}{109} & \multicolumn{ 1}{|c|}{105} &\multicolumn{ 1}{|c|}{104} &\multicolumn{ 1}{|c|}{112} \\
\cline{3-7}
\multicolumn{ 1}{|c|}{} & \multicolumn{ 1}{|c|}{C0} & \multicolumn{ 1}{|c|}{Extra} & \multicolumn{ 1}{|c|}{1.2K} &\multicolumn{ 1}{|c|}{0.6K} &\multicolumn{ 1}{|c|}{3.9K} &\multicolumn{ 1}{|c|}{0.65K}  \\
\cline{2-7}

\multicolumn{ 1}{|c|}{} & \multicolumn{ 1}{|c|}{} & \multicolumn{ 1}{|c|}{Hit} & \multicolumn{ 1}{|c|}{2.45K} &\multicolumn{ 1}{|c|}{2.5K} &\multicolumn{ 1}{|c|}{2.5K} &\multicolumn{ 1}{|c|}{2.63K} \\
\cline{3-7}
\multicolumn{ 1}{|c|}{CK3} & \multicolumn{ 1}{|c|}{C1} & \multicolumn{ 1}{|c|}{Extra} & \multicolumn{ 1}{|c|}{24K} &\multicolumn{ 1}{|c|}{16K} &\multicolumn{ 1}{|c|}{73K} &\multicolumn{ 1}{|c|}{13.5K} \\
\hline
\hline
\end{tabular}
\end{small}
\end{threeparttable}
\vspace{-0.1in}
\end{table}
%\end{scriptsize}
%%%%%%%%%%%%%%%%%%%%%%%%%%%%%%%%%%%%%%%%%%%%%%%%%%%%%%%%%%%%%%%%%%%%%%%%%%%%%%%%%%%
%%%%%%%%%%%%%%%%%%%%%%%%%%%%%%%%%%%%%%%%%%%%%%%%%%%%%%%%%%%%%%%%%%%%%%%%%%%%%%%%%%%
%\begin{scriptsize}
\begin{table} [b]
\vspace{-0.18in}
\centering

\caption{Comparison between \emph{EPIC} and previous works}
\label{MLK-EPIC:benchmark-final-comp}
\begin{threeparttable}
\begin{small}
\begin{tabular}{||ccccccc||}
\hline
\hline
\multicolumn{ 1}{|c|}{Hotspot} & \multicolumn{ 3}{|c|}{C0} & \multicolumn{ 3}{|c|}{C1}\\
\hline
\hline
\multicolumn{ 1}{|c|}{Avg. Perf.} & \multicolumn{ 1}{|c|}{Hit} & \multicolumn{ 1}{|c|}{Extra} & \multicolumn{ 1}{|c|}{Time} & \multicolumn{ 1}{|c|}{Hit} & \multicolumn{ 1}{|c|}{Extra} & \multicolumn{ 1}{|c|}{Time}\\
\hline
\multicolumn{ 1}{|c|}{\emph{EPIC}} & \multicolumn{ 1}{|c|}{92\%} & \multicolumn{ 1}{|c|}{5X} & \multicolumn{ 1}{|c|}{0.72} & \multicolumn{ 1}{|c|}{94\%} & \multicolumn{ 1}{|c|}{4.8X} & \multicolumn{ 1}{|c|}{0.72}\\
\hline
\multicolumn{ 1}{|c|}{ANN\cite{Duo11:aspdac,DuoMLK11:tcad}} & \multicolumn{ 1}{|c|}{89\%} & \multicolumn{ 1}{|c|}{10X} & \multicolumn{ 1}{|c|}{0.3} & \multicolumn{ 1}{|c|}{88\%} & \multicolumn{ 1}{|c|}{8.8X} & \multicolumn{ 1}{|c|}{0.3}\\
\hline
\multicolumn{ 1}{|c|}{SVM\cite{Duo11:aspdac,DuoMLK11:tcad}} & \multicolumn{ 1}{|c|}{86\%} & \multicolumn{ 1}{|c|}{5X} & \multicolumn{ 1}{|c|}{0.35} & \multicolumn{ 1}{|c|}{89\%} & \multicolumn{ 1}{|c|}{5.5X} & \multicolumn{ 1}{|c|}{0.35}\\
\hline
\multicolumn{ 1}{|c|}{PM\cite{A.B.K06:spie,Jingyu07:iccad,Yao06:iccad,Ning08:spie}} & \multicolumn{ 1}{|c|}{85\%} & \multicolumn{ 1}{|c|}{32X} & \multicolumn{ 1}{|c|}{0.2} & \multicolumn{ 1}{|c|}{90\%} & \multicolumn{ 1}{|c|}{25X} & \multicolumn{ 1}{|c|}{0.2}\\
\hline
\hline
\end{tabular}
\end{small}
\begin{scriptsize}
    \begin{tablenotes}
        \item Time calibrated in $hour/mm^2$ unit on 3.2GHz quad-core Linux workstation.
    \end{tablenotes}
\end{scriptsize}
\end{threeparttable}
%\vspace{-0.1in}
\end{table}
%\end{scriptsize}
%%%%%%%%%%%%%%%%%%%%%%%%%%%%%%%%%%%%%%%%%%%%%%%%%%%%%%%%%%%%%%%%%%%%%%%%%%%%%%%%%%%

\begin{figure*}[t]
%\vspace{-0.15in}
 \centering
\includegraphics[width=7.0in]{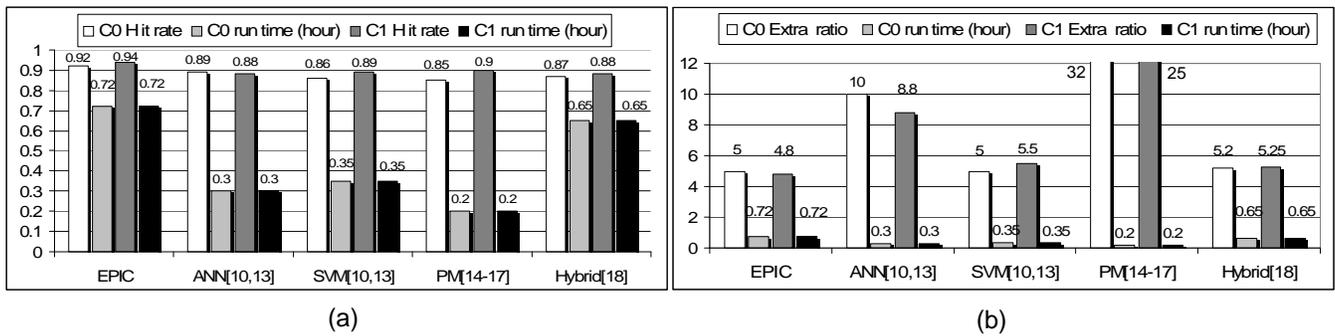}
\vspace{-0.37in}
  \caption{Overall performance comparison in \emph{Hit rate}, \emph{Extra ratio} and run-time over C0 and C1 hotspot data}
  \label{MLK-EPIC:histo}
  \vspace{-0.08in}
\end{figure*}

In Fig.~\ref{MLK-EPIC:histo} we give two summary plots on the performance comparisons of various hotspot prediction methods according to (a): \emph{hotspot accuracy}(\emph{Hit rate})/run-time, and (b): \emph{hotspot false-alarm}(\emph{Extra})/run-time, respectively. Here we make some further observations.

First, \emph{EPIC} achieves much enhanced performance in hotspot prediction accuracy and false-alarms reduction, meanwhile the extra CPU run-time is only 20-30 minutes per $mm^2$ layout in the worst case (versus pattern matching methods).

Second, in comparing C0 category of hotspots with C1 category, we see \emph{EPIC} achieves higher \emph{Hit rate} and lower \emph{False-Alarm ratio} (in the unit of X times of real hotspots) over C1 than C0. In other words, \emph{EPIC} gives more enhancement in accuracy and false-alarm on C1. This is mostly because C1 class represents the set of lithography hotspot with 4.5nm to 6.0nm of EPE, while C0 is the set of hotspots with above 6.0nm EPE values. Under our employed RETs, C0 translates to a set of hotspots that have high variability and small quantity (a few hundred out of a layout with totally millions of patterns); whereas C1 is a set of hotspots with less severe variability and much larger quantity. This is why sometimes detecting C0 could be slightly harder than C1. However it also depends on how the trade-off solution is selected using $\Psi$ based on Fig.~\ref{MLK-EPIC:final-chart} to balance between \emph{hotspot accuracy} and \emph{hotspot false-alarm}, e.g., ANN shows a slightly higher (by 1\%) accuracy rate in C0 than C1 category.

An important advantage of \emph{EPIC} is that it is capable of high performance hotspot prediction under varying EPE thresholds and given large scale design layouts. We see that \emph{EPIC} exhibits very similar CPU run-time when targeting at different categories of hotspot under different EPE thresholds. We have also observed linear run-time complexity when design layout up-scales in area. These properties make our flow efficient for optimizing large industry designs.

Moreover, \emph{EPIC}'s unified formulation covers the static hybrid detection flow proposed in \cite{Jen-Yi11:spie} as a special case, i.e., when \emph{weighting function} $f^{MLK1}(+1)=0.5$ and 0 elsewhere, $f^{MLK2}(+1)=0.5$ and 0 elsewhere, $f^{PM}(+1)=1.0$ and 0 elsewhere, $\theta$ = 1.0, then \emph{EPIC}'s formulation Eqn.(\ref{meta-decision}) will be equivalent to the hybrid flow in \cite{Jen-Yi11:spie}. Here \emph{EPIC}'s advantage lies in the dynamic/automated optimization techniques, thus it can easily reach an optimized solution. Comparing with the static hybrid flow over the employed test cases, \emph{EPIC} observes around 5.7-6.8\% of improvement in \emph{hotspot accuracy} and 3.9-8.6\% of \emph{false-alarm} reduction at a small cost of 10\% extra run-time. Depending on designs, \emph{EPIC}'s advantages could be even higher.

\section{Conclusion}
\label{conc}
In this paper we examined several different types of lithography hotspot detection techniques and proposed EPIC, a new formulation to selectively combine their respective advantages for further accuracy improvement. Under a meta-detection flow, we first used mathematical programming techniques for systematic performance optimization over a set of calibration data, then applied the flow onto new testing cases for further evaluation. EPIC's accuracy, flexibility and false-alarm suppression capability show very promising potential for efficient litho-friendly design. 

}

\scriptsize
\bibliographystyle{unsrt}
\bibliography{MetaMachine_Ref_Long}

\end{document}